\documentclass[12pt,preprint]{aastex}
\newcommand{\be}{\begin{equation}}
\newcommand{\ee}{\end{equation}}
\def\yr{{\rm yr}}
\def\lim{{\rm lim}}

\def\kpc{{\rm kpc}}

\def\kms{{\rm km}\,{\rm s}^{-1}}
\def\masyr{\rm {mas\,yr^{-1}}}
\def\uasyr{\rm {\mu{as}\,yr^{-1}}}
\def\uas{\rm {\mu{as}}}

\shorttitle{Astrometry Beyond the Magnitude Limit}

\begin{document}

\title{Astrometry Survey Missions Beyond the Magnitude Limit}

\author{Samir Salim, Andrew Gould}
\affil{Ohio State University, Department of Astronomy, 140 W.\ 18th Ave., 
Columbus, OH 43210}
\email{samir@astronomy.ohio-state.edu}
\email{gould@astronomy.ohio-state.edu}
\and
\author{Rob P. Olling\altaffilmark{1}}
\affil{Department of the Navy, USNO, 3450 Massachusetts Ave NW, Washington, 
DC 20392-5420}
\email{olling@usno.navy.mil}
\altaffiltext{1}{Universities Space Research Association,
300 D Street, SW, suite 801, Washington, DC 20024-4703}

\begin{abstract}

Three planned astrometry survey satellites, 
{\it FAME}, {\it DIVA}, and {\it GAIA}, all aim at observing
magnitude-limited samples.  We argue that substantial additional scientific
opportunities are within the reach of these mission if they devote
a modest fraction of their catalogs to selected targets that are fainter than
their magnitude limits.  We show that the addition of 
$\mathcal{O}(10^6)$ faint $(R>15)$ targets to the $40\times 10^6$ object 
{\it FAME} catalog can improve the precision of the reference frame by
a factor 2.5, to $7\,\uasyr$, increase Galactocentric distance at which
halo rotation can be precisely ($2\,\kms$) measured by a factor 4, to 25 kpc,
and increase the number of late M dwarfs, L dwarfs, and white dwarfs 
with good parallaxes by an order of magnitude.  In most cases, the
candidate quasars, horizontal branch stars, and dim dwarfs that should
be observed to achieve these aims are not yet known.  We present
various methods to identify candidates from these classes, and assess
the efficiencies of these methods.  The analysis presented here
can be applied to {\it DIVA}
with modest modifications.  Application to {\it GAIA}
should be deferred until the
characteristics of potential targets are better constrained.

\end{abstract}


\section{Introduction}

	The baseline designs of the three planned astrometry satellites
{\it FAME, DIVA,} and {\it GAIA}, call for magnitude-limited samples.
By contrast, {\it Hipparcos} defined only about half of its sample
by a magnitude limit (which ranged from $V<7.3$ to $V<9$), while the 
remaining stars
were chosen based on a variety of specific scientific programs and
objectives.  Does the magnitude-limited approach represent a logical 
choice for future astrometry missions, or does a large fraction of their 
potentially achievable science lie beyond the magnitude limit, as it did for
{\it Hipparcos}?

	Certainly there are strong arguments for choosing most of the
new-satellite samples by magnitude.  For one, these surveys will be
so large, $\sim 10^{7.6}$ stars for {\it FAME} and {\it DIVA} and
$\sim 10^9$ for {\it GAIA}, that almost nothing is known about
this many stars other than their magnitudes and (to some extent) their
colors.  By contrast, {\it Hipparcos} had access to catalogs of
high proper-motion, nearby, and other classes of special stars
whose sizes were a substantial fraction of its much smaller $\sim 10^5$
star catalog.  Also, the astrometric precision of the new missions
falls off rapidly with flux, either $\sigma\propto \rm flux^{-1/2}$ in the
photon-noise regime or $\sigma\propto \rm flux^{-1}$ in the read-noise regime.
Thus it seems logical to focus the effort on the stars for which the
astrometry is best.  This effect was much less compelling in the case
of {\it Hipparcos} for which the astrometric precision fell off only
by a factor $\sim 2$ between $V=7.3$ and $V=11.5$ (because it was able
to devote longer exposures to fainter objects).  Finally, if
{\it Hipparcos} had established its full catalog simply by setting a 
magnitude limit
(at about $V\sim 9$), whole classes of stars would have been virtually
absent, including M dwarfs, white dwarfs, and radio stars that link the
optical and radio frames.   The new missions will go so much deeper that they
will contain substantial samples of all of these various objects, even
without any special effort.

	Although these arguments have some merit, we believe that
if the new missions maintain their purely magnitude-limited approach,
they will miss out on major scientific opportunities.  Some of these
losses are obvious.  For example, dim stars (like white dwarfs and M
dwarfs)
are relatively quite close even when they lie below the magnitude limit,
and hence they can have relative parallax errors that are very
small even though their absolute astrometry is poor.  But as we will
show, there are other, less obvious faint objects that can have huge
scientific returns despite the fact that they are so far away that their
{\it parallaxes} cannot be even crudely measured.  Moreover, once one 
recongizes
the importance of observing these various categories of faint objects,
it is far from obvious how to assemble the samples, which are likely
to contain $\mathcal{O}(10^6)$ stars, most of which are not known to
be in the designated classes, or are not yet identified.

	Here we examine what general classes of faint stars should
be considered as potential observing targets by astrometric missions,
despite the fact that in some cases, each faint star so chosen may displace
some bright star which would have yielded better astrometry.  We review
how {\it Hipparcos} handled these classes and then look forward to
future missions.  We identify a few major science questions that can
be attacked if {\it FAME} or {\it DIVA} chooses to observe faint stars.
We make quantitative estimates of how well {\it FAME} will be able
to address these questions with and without the observation of faint
stars and assess various approaches to constructing an input catalog
which {\it FAME} requires.
The principles that we outline can be applied directly to {\it DIVA}, 
so we do not repeat the analysis for it.  
We briefly discuss how these principles
will apply to {\it GAIA}, but argue that a detailed analysis is not warranted 
until closer to mission launch.

	We do not believe that we have exhausted the scope of interesting
faint stars that could be added to future mission catalogs.  
Our goal is rather to present
a few classes of additional stars that could have major scientific
impact and to illustrate both the promise and the difficulties of
including them.  We hope that this will encourage others to explore
the potential of other categories of stars and to improve on the methods
that we outline here for selecting the candidates.

	In the next subsection we analyze target selection for {\it Hipparcos}
catalog in terms of magnitude-limited vs.\ faint sample. In \S\ 1.3 we
introduce the methodology of non-magnitude selected samples to future
astrometric missions. These missions are described in general in \S\ 1.2,
while their astrometric performance is evaluated in \S\ 2. In \S\S\ 3-6 we
apply this methodology to {\it FAME}, as a specific example\footnote{After
submitting the manuscript, the {\it FAME} mission, at least in the form we
present it here, has been cancelled by NASA. Efforts are presently being made
to continue the project. In any case the methodology, the proposed scientific
objectives, and the target selection methods are applicable to astrometry
missions in general.}. In \S\ 7 we discuss how this methodology can be applied
to {\it GAIA} and {\it DIVA}.

\subsection{The {\it Hipparcos} Experience}

\begin{figure}
\plotone{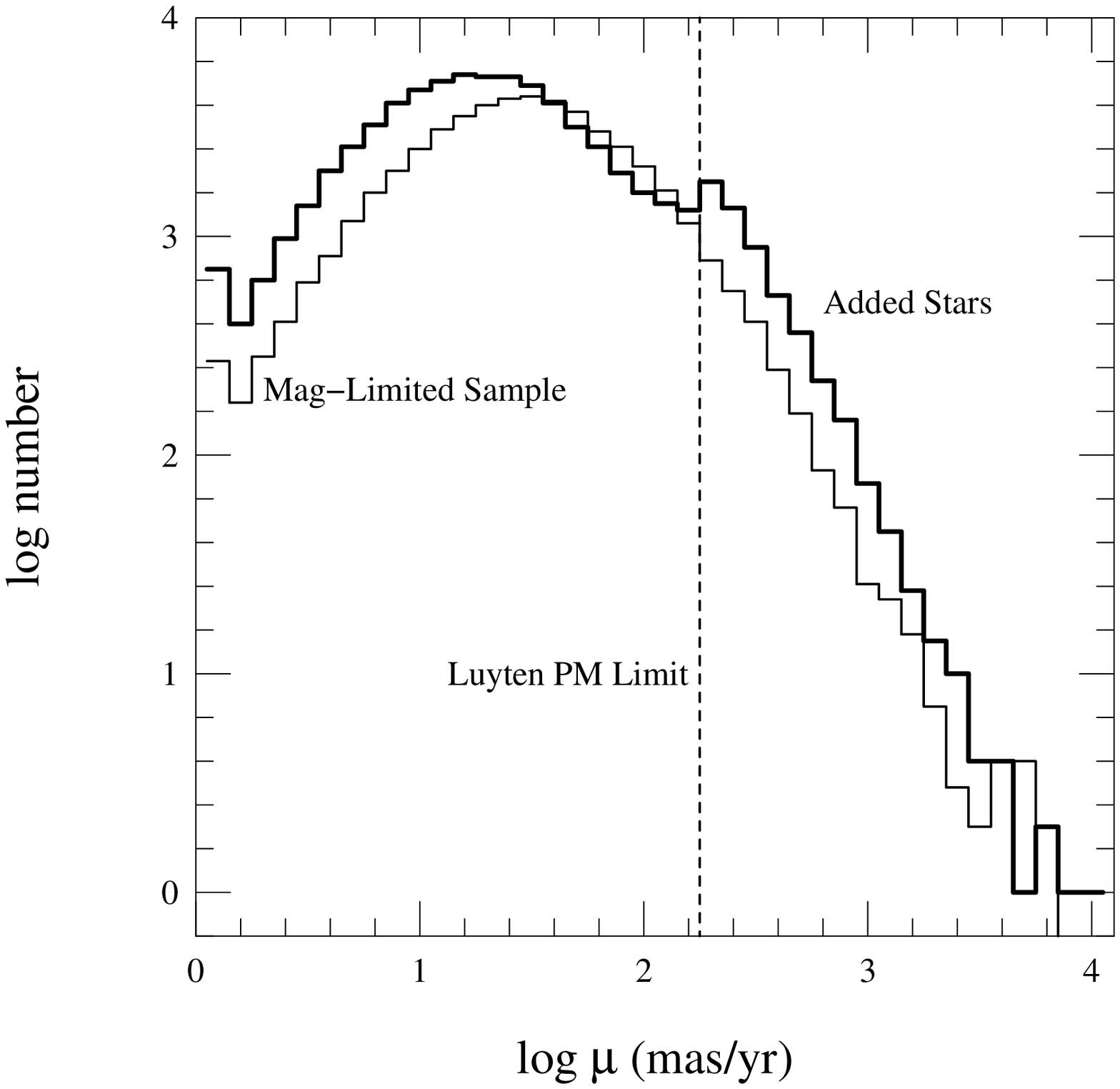}
\caption{\footnotesize Number of {\it Hipparcos} stars lying within
({\it solid histogram}) or beyond ({\it bold histogram}) 
the {\it Hipparcos} magnitude limit as a function of their measured
proper motion, $\mu$.  (This mag limit ranges from $V=7.3$ to $V=9.0$,
depending on Galactic latitude and spectral type.)\ \  
The effect of adding stars drawn from the \citet{nltt} NLTT catalog
with its proper-motion limit of $\mu=180\,\masyr$ ({\it dashed line})
is clearly visible.  
Many of these stars are dim (and hence close) and therefore
have good parallaxes.  However, the majority of added stars have
very small ($\sim 15\,\masyr$) proper motions, and hence barely
measurable parallaxes.  Like {\it Hipparcos}, future astrometry missions can 
profit by adding both classes of stars to their otherwise magnitude-limited
catalogs.
\label{fig:hipp}}
\end{figure}

The magnitude-limited component of the {\it Hipparcos} catalog
comprises roughly 50,000 stars that satisfy $V<7.3+1.1\sin|b|+\Delta V$,
where $\Delta V=0.6$ for spectral types G5 and earlier and is zero otherwise.
The remaining roughly 70,000 stars (with $7.3<V\la 11.5$) were chosen from 
among 214,000 submitted to ESA in 214 separate proposals.  Neither the
proposals themselves, which supported science projects ranging from the 
sub-dwarf distance scale to improvement of the
lunar orbit, nor the process of selection
can be properly reviewed here.  The most important points we want to
make can be understood by inspection of Figure \ref{fig:hipp}, which shows the
proper-motion distributions of {\it Hipparcos} stars that lie respectively
within or beyond the
magnitude limit.  The first point to note is that the great majority 
($\sim 70\%$) of the stars with
proper motions exceeding the $\mu=180\,\rm mas\,yr^{-1}$ limit of the 
Luyten (1979, 1980) NLTT catalog are drawn from beyond the magnitude limit.
This is because {\it Hipparcos} included essentially all NLTT stars
that lie within its range $(V\la 11.5)$.  Over 99\% of the roughly 1000
intrinsically dim ($M_V>8.5$) stars in the {\it Hipparcos} catalog
lie beyond its magnitude limit, and of these almost 90\% come from the
NLTT.  Thus, while various additional sources of dim stars
were exploited (like nearby stars), proper-motion selection was the most 
effective method for including them.

	The second point to note, however, 
is that the great majority of the added
stars have very low proper motions, $\mu\sim 15\,\rm mas\,yr^{-1}$,
meaning that they probably lie at distances of order 500 pc, which implies
that their parallaxes could hardly be measured by {\it Hipparcos}.  The science
drivers for these stars are varied, but in most cases knowledge of their
distances was not critical.  For example, for stars that were to be
occulted by the Moon, only the positions and proper motions would be
of interest.  Similarly, RR Lyrae star proper motions 
would be useful for a statistical parallax measurement even though their trig
parallaxes were marginal at best.  Another 200 stars were chosen because
they lay close to quasars, the hope being to eventually use these to
tie together the {\it Hipparcos} and extragalactic reference frames.  
Again, no parallaxes are needed to make this tie-in.

	Thus, the main lesson from {\it Hipparcos} is that there can
be a wide variety of reasons for extending the catalog beyond the
magnitude limit.  In some cases, this may be the only way to obtain good
parallaxes of special classes of stars, but in other cases the parallaxes
may be of no interest.

\subsection{Future Missions}

Four major astrometry missions are planned for the next decade,
{\it FAME}, {\it DIVA}, {\it GAIA} and {\it SIM}.  Three of these exploit 
the same basic principle of the scanning telescope, employed first by 
{\it Hipparcos}. 
{\it FAME}\footnote{{\it Full-Sky Astrometric Explorer}:
\url{http://www.usno.navy.mil/FAME/}} \citep{fame}
is a NASA mission proposed by United States Naval Observatory (USNO). 
  {\it FAME}'s
design specifications promise twenty-fold improvement in astrometric precision
over {\it Hipparcos} in the $5<R<10$ range, down to 0.5 mas precision at the
survey limit of $R=15$. Its aim is to obtain astrometry, as well as photometry
in two Sloan bands ($r'$ and $i'$), of $\sim4\times10^7$ stars, over a 5 yr
planned mission duration. The German-built satellite 
{\it DIVA}\footnote{\url{http://www.ari.uni-heidelberg.de/diva/}}
is now near final approval.  It is expected to observe a similar number of 
stars, but only for 2 years and with somewhat smaller mirrors.  Hence it should
achieve several times less precise astrometry than {\it FAME}. In addition, 
{\it DIVA} expects to obtain 30-element spectral photometry.  Both
{\it FAME} and {\it DIVA} aim to launch in 2004.

The much more ambitioius {\it GAIA}\footnote{\url{
http://astro.estec.esa.nl/GAIA/}} mission
will obtain an order of magnitude better performance than {\it FAME}. It 
will observe $\mathcal{O}(10^9)$ objects to $V=20$ including not only astrometry,
but photometry and radial velocities as well.  However, {\it GAIA}'s 
earliest launch date is 2011.  Also planned for this epoch
is the {\it Space Interferometry Mission (SIM)}, which will
be yet more precise.  However, since {\it SIM} is a targeted rather than
a survey mission, it does not fall within the framework of the present study.

\subsection{Scope and Method}

	Following the experience of {\it Hipparcos}, there are two
broad categories of potential additional targets that lie beyond the 
magnitude limit:
those that are so dim (and hence so close) that one could obtain useful
parallaxes even with relatively poor astrometry, and those for which
the positions and proper motions are of interest even when the parallaxes
may be unmeasurable.  While these lessons are very general, their
application is not.  When the flux limit falls by a factor 100 or 1000,
as it does when going from {\it Hipparcos} to {\it FAME} or from
{\it FAME} to {\it GAIA}, the very classes of objects that fit into
these two broad categories can change completely.  The problems in
identifying members of these classes change completely as well.
For example, {\it Hipparcos} observed 10,000 stars that are occulted
by the Moon, the majority of which were beyond the magnitude limit.
{\it FAME}, by contrast, will observe about 3 million such stars
{\it within} 
its magnitude limit, far more than could be used in any conceivable 
lunar-orbit or lunar-topography investigation.  On the other hand, consider
quasars, a class of objects whose proper motions are of interest because they
are known a priori to be zero.  {\it Hipparcos} observed precisely one
of these (3C 273), and the proper-motion error of this measurement,
$\sigma_\mu \sim 6\,\masyr$, turned out to be so large as to render
it useless.  The {\it Hipparcos} reference frame was in
fact fixed to radio stars, not directly to quasars.  By contrast,
as we discuss in detail in \S\ 3.2, the {\it FAME} reference frame
can be fixed with exquisite precision by observing $\mathcal{O}(10^5)$ quasars,
the vast majority of which lie beyond the magnitude limit, and indeed
are yet to be discovered.

	Thus, our approach will be to apply the very general lessons from
{\it Hipparcos} to the concrete situation of {\it FAME}.  We will address
both the questions of what science can be obtained by going beyond the
magnitude limit, and of how one can find the often unclassified targets.
The characteristics of the {\it DIVA} mission are similar enough to
those of {\it FAME} that the range of possible science projects
will be similar.  As we discuss
in \S\ 7, the science achievable beyond {\it GAIA}'s magnitude limit
can only be sketched at the present time.

\section{Astrometric Performance of Survey Missions}

Generically, one expect three regimes of precision for an astrometric
satellite: systematics limited, photon limited, and read-noise limited, for
bright, middle, and faint stars respectively. Here, our focus is on the latter
two regimes, where the astrometric errors should scale as $\sigma
\propto {\rm flux}^{-1/2}$ and $\sigma \propto {\rm flux}^{-1}$,
respectively. The one-dimensional positional accuracy of an object after a
mission of duration $t$ is then given by an expression of the form, 
\be
\label{eqn:sens}
\sigma = \sigma_0 \sqrt{5\, \yr \over t} 10^{0.2(m-15)}[1+c_{\rm
RN}10^{0.4(m-15)}]^{1/2}, 
\ee 
where $m$ is the astrometric-band magnitude, and 
where we have scaled the mission duration to 5 yr, and the magnitude to 
$m=15$. Equation (\ref{eqn:sens}) contains two constants that need to be 
determined: the normalization factor $\sigma_0$, and the read-out noise 
coefficient $c_{\rm RN}$. In fact, since all future missions are planning
to use an {\it Hipparcos}-like scanning law, which has highly uneven
coverage, the ``constant'' $\sigma_0$ is actually a function of ecliptic
latitude.  For statistical studies of the type investigated here, it
is appropriate to take an average value.

Proper-motion accuracies for equally spaced observations are related to
positional accuracies by 
\be
\label{eqn:mu}
\sigma_{\mu} = \sqrt{12}{\sigma\over t}
\ee 
While {\it Hipparcos}-like observations are not precisely equally spaced,
this formula remains accurate for mission lifetimes of at least 2 yr.

Parallax accuracies depend on the ecliptic latitude of the
object, and range from $\sigma_{\pi} = \sigma$ at the ecliptic poles to
$\sigma_{\pi} = \sqrt{2} \sigma$ on the ecliptic, again assuming uniform
sampling. Since there will be some times of the year when no observations are
carried out, the actual parallax precision will not be given as simply, and
will be somewhat worse.  For {\it FAME},
\citet{mur} uses a realistic scanning law and finds 
that on average, 
\be
\label{eqn:sigmapi}
\sigma_\pi \simeq 1.43 \sigma,
\ee
 (see also Olling 2001), 
which  we will use here.  This coefficient should depend only very weakly
on the precise mission characteristics.

\subsection{Specific Estimates}

For the three proposed missions, the parameters entering equations
(\ref{eqn:sens})--(\ref{eqn:sigmapi}) are approximately given by
\be
\label{eqn:cons}
\sigma_0 = 240\, {\uas}, \quad c_{\rm RN}=1.06,\quad m=R,\quad t=5\,{\rm yr}. 
\quad {\it (FAME)}
\ee  

\be
\label{eqn:consdiva}
\sigma_0 = 965\, {\uas}, \quad c_{\rm RN}=1.91,\quad m=R,\quad t=2\,{\rm yr}. 
\quad {\it (DIVA)}
\ee  

\be
\label{eqn:consgaia}
\sigma_0 = 1.8\, {\uas}, \quad c_{\rm RN}=0.012,\quad m=V,\quad t=5\,{\rm yr}. 
\quad {\it (GAIA)}
\ee  

Since in all cases the actual astrometric bands differ from the standard
bands in these equations, the parameters vary slightly with stellar
type.  For example, for {\it FAME} $|R-m|$ is typically small,
but does rise to $R-m$ = 0.27 for an M7 dwarf.  Throughout this paper we 
refer to $R$ magnitudes (and not $V$, for example) to describe {\it FAME}
sensitivity. In our actual calculations we take account of the small 
differences between the {\it FAME} astrometric band and $R$ as a function
of spectral type, although in practice this makes very little difference.
Corrections to convert $V$ to $m$ for stars of different 
temperatures are given in \citet{oll}.

	From equations (\ref{eqn:cons})--(\ref{eqn:consgaia}), the
three missions reach the read-noise limit (where photon noise equals read 
noise) at $R=15.0$, $R=14.3$, and $V=19.8$.  These are also approximately
the magnitude limits of these surveys ($R\sim 15$, $R\sim 15$, and $V\sim 20$).
Nevertheless, while the astrometric precision obviously deteriorates rapidly
beyond these limits, the scientific potential from observing carefully
chosen fainter objects is high.  In the following three sections we illustrate
this point, specifically using the {\it FAME} parameters 
(eq.\ [\ref{eqn:cons}]).  In \S~7, we briefly discuss {\it DIVA} and 
{\it GAIA}.

\section{Quasar Reference Frame}

In order to be able to translate relative proper motions to absolute ones, the
reference frame must be ``anchored'' to a quasi-inertial frame: the satellite
must observe a number of objects that have essentially no proper motion, and
whose light profile is stellar, or objects that move but whose absolute proper
motion is known with great precision. With the roughly 100 quasars that fall
within {{\it FAME}'s} survey limits, the frame can be fixed only with a
precision of $\approx 19\,\uasyr$ (in one direction).

Further refinement of the rotation of the {\it FAME} frame is possible by
including in the input catalog quasars (or quasar candidates) with $R>15$. The
precise fixing of the reference frame is important as it allows the motion of
certain populations of stars to be determined with much greater accuracy,
especially increasing the fractional accuracy for slow-moving
populations. Such measurements will be described in more detail in the \S\S
4.1. and 4.4. Of perhaps even greater general importance, measuring the 
`parallaxes' of objects that actually have no measurable parallaxes, such as
quasars, offers a unique check on any unmodeled or unknown systematics that
might affect the spacecraft while taking data, or anywhere later in the data
reduction process.

The precision of the frame fixing depends on the number, distribution on the
sky, and apparent brightness of the quasars that {\it FAME} will observe. Let
$\hat{\mathbf n}_k = (\hat{n}_1,\hat{n}_2,\hat{n}_3)_k$, the unit vector
toward the $k$-th quasar, be given in the rectangular Galactic coordinates,
with the third component ($z$) perpendicular to the Galactic plane. We then
form rank-3 matrix $\mathbf{b}$ 
\be
\label{eqn:qso}
b_{ij} = \sum_{k=1}^n \left({\delta_{ij} - \hat{n}_i\hat{n}_j \over
\sigma_{\mu}^2}\right)_k; \qquad i,j=1,2,3 
\ee 
where $\sigma_{\mu}$ is given
by equation (\ref{eqn:mu}) and depends on the quasar's apparent magnitude, and
$\delta_{ij}$ is the Kronecker delta. Then the reference frame accuracies
along $x$, $y$, and $z$ axes are given as square root of the diagonal terms of
the covariance matrix ${\it {\bf c}} = {\it {\bf b}}^{-1}$.

\subsection {Quasar (Candidate) Selection}

In order to implement the frame fixing, one must specify how to build the
quasar sample that {\it FAME} will observe and what is the number of quasars
or quasar candidates in this list. The number of objects is important because
we want to maximize the scientific gain (precision of the frame) without
taking up too great a part of {\it FAME} input catalog.

Ideally, one would like to include {\it all} quasars in the sky down to some
limiting magnitude. Obviously, with a 100\% complete survey we always gain
maximum astrometric signal with the least number of objects. Unfortunately the
currently known sample of quasars (\citealt{veron}, hereafter VV00) is complete
only to $R\approx 14.5$. Completeness drops to about 50\% only a magnitude
fainter, and is less than 10\% at $R\approx 18$. (The completeness was
assessed with respect to QSO sky densities as given by \citealt{hs}. Here, and
in the rest of the paper we use mean quasar colors of $B-V=0.25$ and
$V-R=0.25$ to conveniently convert between $B$ magnitudes used in \citealt{hs},
the mostly $V$ magnitudes used in VV00, and our $R$ magnitudes.)

Fortunately, there are several surveys underway that will discover new quasars
over wide areas of sky. For definiteness, we will assess what
quasars are likely to be known
in late 2003. We stress that the
``quasar'' list can actually be comprised of quasar candidates, i.e.\ no
spectroscopic confirmation is required at the time of catalog compilation.  It
is only when the data are reduced at the end of the mission that spurious
objects must be eliminated so that their systematic motion does not corrupt
the frame. For quasar candidates that will not have their spectra taken, {{\it
FAME}'s} measurements of proper motion will serve as a criterion to eliminate
white dwarf contamination. For blue horizontal branch stars which are much
farther away than the white dwarfs and have typical proper motion on the order
of the measurement accuracy, the difference $\Delta(u'-g')\approx1.0$ of color
indices between them and quasars can be used if $u'$ photometry is
available. Otherwise, quasars can be distinguished from blue stars by their
$K$-band excess. Quasars are also indicated by their variability which is
different from that of RR Lyr stars. Eventually, the fractional
contamination must be $\la N_{\rm QSO}^{-1/2}$, where $N_{\rm QSO}$ is the 
number of quasars in the sample, so that the systematic errors are smaller
than the statistical ones. For any quasars candidates
that lack spectroscopic confirmation and are therefore potential contaminants,
one can decrease their effects on frame degradation by assigning them
different weights in the fit, based on the probability of them being quasars
constrained from all available (radio, optical, IR, UV, and X-ray)
information.

One survey that might produce an {\it all-sky} sample of quasars virtually
complete to beyond $R=19$ is the {\it Galaxy Evolution Explorer (GALEX)}. {\it
GALEX} is an imaging and spectroscopy space mission operating in the UV (130-300 nm)
region. It will perform slitless spectroscopy of $10^5$ galaxies and $10^4$
quasars and obtain all-sky two-band UV photometry of $10^7$ galaxies and $10^6$
quasars. The two-band photometry itself will be enough to confirm the AGN-like
spectral energy distribution, and thus distinguish quasars from stellar
objects (mostly white dwarfs, and A-colored halo stars) (B. Peterson 2001,
private communication). With the launch scheduled for January 2002, the
results of the all-sky {\it GALEX} survey might be complete approximately at
the time when the {\it FAME} input catalog is to be finalized (C. Martin 2001,
private communication), but at the moment that remains a major uncertainty.

Hence it is prudent to determine what can be done in the absence of {\it
GALEX} data. We expect most of the good quasar candidates to come from the
Sloan Digital Sky Survey \citep[SDSS]{sdss}. SDSS will also be essentially
complete to beyond $R=19$, and will perform photometry in five bands which
should be very effective (70\% or higher, Richards et al.\ 2001) in selecting
the quasar candidates. More importantly, it will provide spectroscopic
confirmation for almost all of the bright (by SDSS standards) candidates that
are relevant for {\it FAME}. However, SDSS is not an all-sky survey ---it
covers the north polar cap ($b\la 30\degr$), and 3 strips totaling 700 ${\rm
deg}^2$ in the south polar cap, and therefore it cannot substitute for {\it
GALEX}. Again, the issue of the status of SDSS and the availability of its
data at the time required for {\it FAME} is important. The current estimate
(D. Weinberg 2001, private communication) is that 1/2 of the northern cap and
all of the southern strips will be available.

The observing list will be further augmented by quasars found in two southern
strips (750 ${\rm deg}^2$ total) that make the 2dF Quasar Survey
\citep[2QZ]{2qz}, where we expect to get all quasars with
$R>16.5$. Another wide-coverage sky survey being undertaken is the FIRST
Bright Quasar Survey \citep[FBQS]{fbqs}, in which quasars are selected by
matching pointlike radio sources from FIRST to stellar-like objects with
quasar colors found in Schmidt optical survey plates. The FIRST sky coverage
coincides with that of SDSS so we might expect some additional candidates from
the half of the northern cap that will not be available from SDSS in late
2003. The detection efficiency of FBQS is comparatively low ($\sim 25\%$), so
it cannot fully substitute for the lack of SDSS coverage.

We also consider the possibility of selecting quasar candidates in regions
that will not be covered by SDSS or 2QZ (50\% of the northern ($b>30\degr$)
cap and 85\% of the southern ($b<-30\degr$) cap.) To this end we investigate
using the Two-Micron All-Sky Survey (2MASS) Point Source Catalog
(PSC). \citet{bark} have shown, using the 2nd Incremental Release of 2MASS PSC
(covering $\sim 50\%$ of the whole sky), that 2MASS counterparts of the VV00
quasars occupy a distinct part in the ($B_{\rm USNO}-J$) vs. ($J-K_s$)
color-color diagram. The ($J-K_s$) color of matched quasars is redder
than that of stars, while ($B_{\rm USNO}-J$) color is
moderately blue compared to the stellar range (both $B_{\rm USNO}$ and 
$R_{\rm USNO}$ magnitudes in 2MASS come from USNO-A2.0 catalog \citep{usnoa2}, which is 
based on first generation sky survey plates). How efficient can 2MASS
potentially be in detecting quasars to some apparent magnitude?  In order to
determine this, we simply compared the number of 2MASS-matched VV00 quasars to
a complete VV00 sample (corrected for 2MASS sky coverage) in different
magnitude bins. We find that 2MASS is basically 100\% complete for $V<17.5$, 
and its completeness drops to 50\% around $V\sim18.5$. Therefore, 2MASS would appear to be a good place to find quasars. The main problem, however, is that even when
selecting only 2MASS objects away from the Galactic plane that have colors
typical of quasars and are away from the stellar locus, and have proper
motions (inferred from the distance between 2MASS objects and their USNO A2.0
counterpart) consistent with zero, the stellar contamination is still
quite high. We will return to this problem shortly.

\subsection{Reference Frame Accuracy}

In order to calculate the frame precision (except for VV00) we simulate
observations from a mock $R>14.5$ quasar catalog with an apparent magnitude
distribution according to \citet{hs}, and including a simple model of Galactic
extinction. For the bright ($R\leqslant 14.5$) end of quasars we use the
actual VV00, since we found it to be complete in that magnitude range. In
fact, in this bright part, we remove objects from VV00 that do not have a
stellar appearance, although {\it FAME} could possibly make good measurements
of some non-pointlike AGN that exhibit significant nucleus-host
contrast. Intrinsic photocenter variation is ignored since it is considerably
smaller than the astrometric precision for the great majority of quasars.

\begin{figure}
\plotone{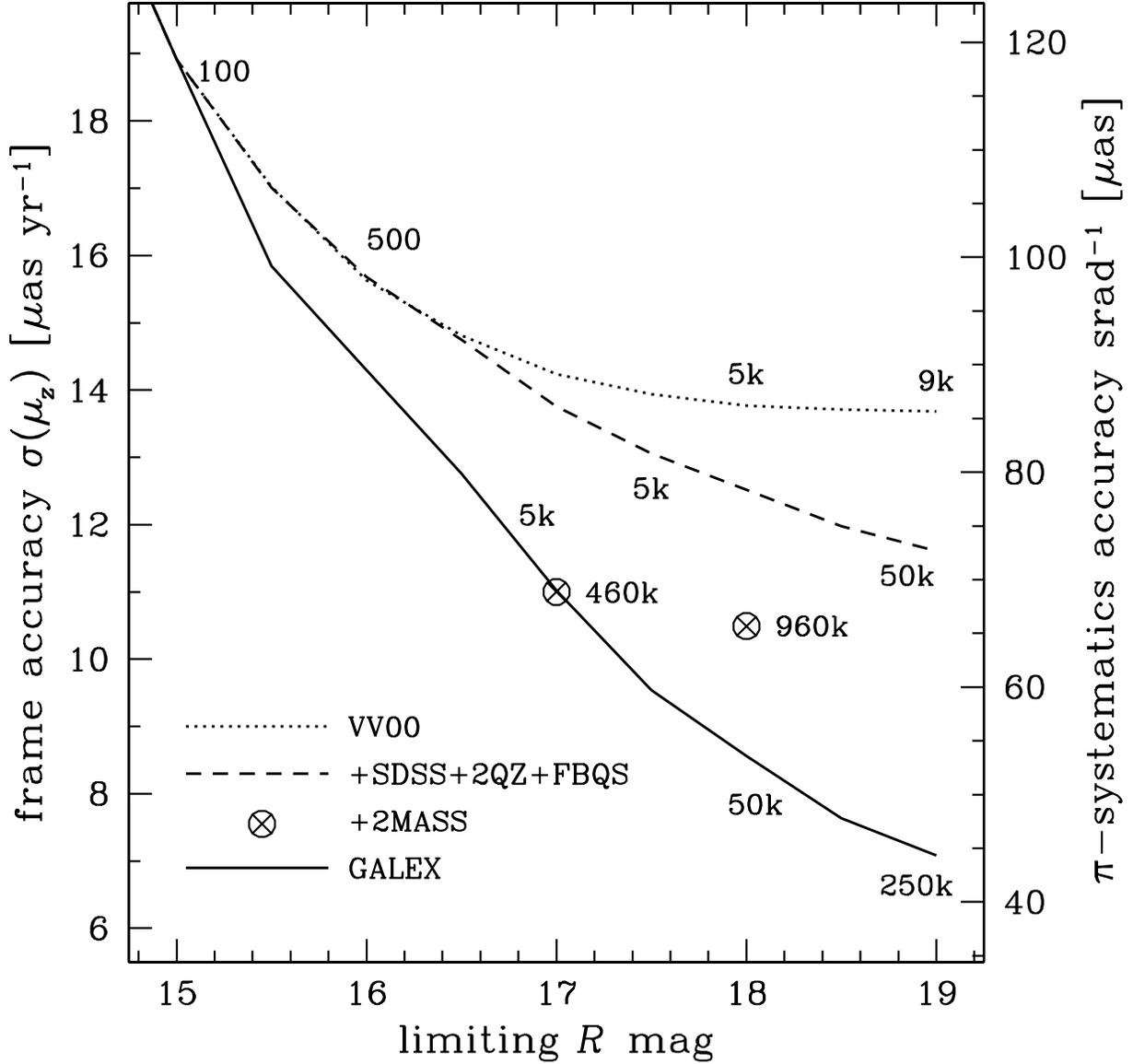}
\caption{\footnotesize Left-hand axis: Precision of reference frame obtained by observing quasars, assuming {\it FAME} sensitivities. The three lines show how the accuracy of the frame improves when adding more quasars to some limiting magnitude $R$. The uppermost line corresponds to the accuracy achieved by observing the currently known quasars \citep[VV00]{veron}. The middle line shows the accuracy when one adds quasars to be identified by SDSS, 2QZ and FBQS surveys by the end of 2003, in time to be included in the input catalog. The best accuracy, lower line, is achieved if faint quasars are found all over the sky by a mission like {\it GALEX}. In the possible absence of {\it GALEX} more quasars can be found using 2MASS, the corresponding accuracy shown with circled crosses. Next to the lines and points we note the number of quasar candidates that must be put into the input catalog. Right-hand axis: Observing quasars also allows parallax systematics to be checked. This ac!
curacy is shown normalized to 1 srad of the sky. \label{fig:qso}}
\end{figure}

In Figure \ref{fig:qso}, on the left-hand axis, we show the accuracy of the
reference frame achieved in the $z$-direction if different surveys are
included and observed by {\it FAME} to some limiting $R$ magnitude. Next to
each line we place labels indicating the approximate number of objects
brighter than that point. We use equation (\ref{eqn:qso}) to calculate the
frame accuracy. The short-dashed line shows what frame accuracy is achieved if
only VV00 quasars are used. This is roughly equivalent to asking what quasars
{\it FAME} would be able to observe if it flew now.  One can see that the
accuracy does not improve much beyond $R=17$. The VV00 sample includes 1800
quasars to $R=17$ and produces $\sigma_z = 14.2\, \uasyr$, while including
5000 quasars to $R=18$ gives only a slightly better accuracy. Better results
are obtained when we include the quasars that SDSS will identify by late 2003,
plus those from the 2QZ and FBQS surveys. They are shown with the long-dashed
line. In this calculation, for regions not covered or inefficiently covered by
SDSS, 2QZ and FBQS we keep the VV00 quasars. With these surveys {\it FAME} can
get progressively better result as fainter and fainter quasar candidates are
included. At $R=17$, with 2500 quasar candidates (and assuming they are all
actually quasars) one could get $\sigma_z = 13.8\, \uasyr$. If the systematics
allow photometry of $R=19$ objects (71,000 quasars total), the achieved
precision becomes $11.6 \, \uasyr$. The full line, which we can also consider
an ideal case, shows how the quasars detected by {\it GALEX} could constrain
the frame accuracy. We exclude from the calculation {\it GALEX} quasars within
$10\degr$ of the Galactic plane, as these will be difficult to distinguish by
{\it FAME}. At $R=17$, with only 5100 quasars, {\it FAME} already reaches
$\sigma_z = 11.0\, \uasyr$, which improves to $8.6 \, \uasyr$ at $R=18$ and
40,000 quasars, and could ultimately reach $7.1 \, \uasyr$ at $R=19$, using
some 270,000 quasars.

As mentioned earlier, one can in principle improve over the situation caused
by the possible absence of {\it GALEX} data and the incomplete sky coverage of
SDSS to be available in late 2003 by including quasar candidates selected
using 2MASS. However, due to non-negligible stellar contamination in color and
proper-motion selected objects from 2MASS, improvements in the frame accuracy
are possible only at the cost of including many more objects in the input
catalog. In Figure \ref{fig:qso} we show with two circled crosses the
improvement in frame accuracy by including in the input catalog quasar
candidates selected from 2MASS that have $R_{\rm USNO}$ magnitudes less than
17.0 and 18.0 respectively, lie in north and south polar caps with the radius 
of $60
\degr$, centered $10\degr$ away from the Galactic poles (in order
to reduce stellar contamination from the bulge), have proper motions
consistent with zero, and satisfy the color selection criteria
\be 0.2<B_{\rm USNO}-J\leqslant1.8; \qquad 
J-K_s\geqslant 0.6+0.2(B_{\rm USNO}-J).  
\ee 
This color selection is chosen to maximize quasar detections ($\sim 70\%$
detection efficiency), while keeping the number of stellar contaminants as
small as possible. We estimate that $\sim 2\%$ of the candidates selected in 
this way are actually quasars.

We find that by including 460,000 2MASS candidates to $R_{\rm USNO}=17$ one can
achieve frame accuracy of $\sigma_z = 11.0\, \uasyr$, and with 960,000 
candidates to $R_{\rm USNO}=18$ reach $\sigma_z = 10.5\, \uasyr$. It is
possible that by adopting some different color criteria, or by avoiding parts
of the sky with particularly high stellar contamination, the improvements
similar to these can be achieved at a somewhat smaller cost in terms of number
of candidates.

Frame accuracy in the $x$ and $y$ directions is 10\% better than in the $z$
direction for the {\it GALEX} sample, regardless of magnitude. For all other
surveys the $x$ and $y$ accuracy is 13\% better at $R=16$ increasing linearly
to 20\% at $R=19$. Accuracy in the $z$ direction is in all cases the worst,
 because there are fewer quasars close to the Galactic plane.

We previously mentioned that measuring quasar positions can serve as a check
on the systematics of the satellite accuracy. Say, for example, that there is
some unmodeled annual effect that changes the basic angle between the two
fields of view of {\it FAME}. This would cause false parallaxes in the certain
regions of the sky. One can independently check this is by observing objects
with no parallax, like quasars. On the right-hand side axis of Figure
\ref{fig:qso} we note values of the accuracy of parallax systematics in 1 steradian
of the sky obtainable with different quasar samples. (We assume that for the
random direction positional accuracy is approximately proportional to
$z$-direction proper motion accuracy. See previous paragraph.) The precision
(45 - 75 $\uas$) is adequate to allow detection of systematic effects that are
of the order of the mission's best statistical errors.

In addition to astrometry, observing quasars with continuous photometric
sampling at an average rate of 1 day$^{-1}$, will permit variability studies
with 12\% precision per observation at $R=16$.

\section{Kinematics of the Galaxy}

\subsection{Proper Motion of the Galactic Center and the Motion of the LSR}

The proper motion of the Galactic center is the reflex of the Sun's motion 
around
it. Assuming that the compact radio source Sgr A* is at rest with respect to
the dynamic center of the Galaxy, which the huge mass of a black hole
associated with it suggests, radio astrometry of Sgr A* by \citet{backer} and
\citet{reid} yields $\mu_{\rm GC} = -6.2\pm0.2\,\masyr$ and $\mu_{\rm GC} =
-5.9\pm0.4\,\masyr$, respectively. While not discrepant, these two
measurements imply that the proper motion is known to only $5\%$. By measuring
proper motions of great number of bulge stars that are part of the main survey
mission ($R<15$), the statistical error of the mean proper motion will almost
vanish, and the ensuing precision of the proper motion of the Galactic center
will be limited by the precision with which the reference frame rotation is
known. In the previous section we described scenarios in which {\it FAME}
would be able to determine the frame rotation to, say, $7\,\uasyr$. This would
therefore allow the proper motion of the Galactic center to be determined to 
$0.1\%$, a fifty-fold improvement over the current situation.

Since $\mu_{\rm GC} = V/R_0$, where the velocity $V=\Theta_0+V_\sun$ is the
sum of the rotation speed of the Local Standard of Rest (LSR) and Sun's motion
with respect to it, and $R_0$ is the distance to the Galactic center, the
proper motion constrains the ratio of these quantities. Since the {\it FAME}
measurements will practically eliminate statistical uncertainty in $\mu_{\rm
GC}$ and $V_\sun$, and the uncertainty in $R_0$ which is currently around
$5\%$ might eventually go below $1\%$ \citep{R_0}, {\it FAME} would be able to
determine $\Theta_0$ with $2\,\kms$ accuracy. Measuring the value of the Milky
Way's rotation curve at the Solar circle with such accuracy will improve estimates
of its dynamical mass. One will also be able to put constraints on the
possible motion of LSR perpendicular to the Galactic plane. Such motion might
be the result of the triaxiality of the dark halo. \citet{binney} predicts
$\mu_{{\it GC},\perp}=27\,\masyr$, assuming ellipticity $\epsilon=0.07$. While
today's measurements cannot measure this effect at all, with the frame fixed
by faint quasars, {\it FAME} would be able to measure it with 25\%
accuracy. Other types of non-axisymmetry, like the Galactic bar, could also be
explored to better accuracy in this way.

If the radio measurements of Sgr A* eventually reach comparable precision, it 
will be possible to test with high precision the premise that Sgr A* 
represents the dynamical center of the Galaxy by comparing the Sgr A* proper 
motion to the mean proper-motion of bulge stars measured by {\it FAME}. 
In the direction perpendicular to the Galactic plane, this test would be 
essentially free of possible systematic effects.

\subsection{Rotation of the Galactic Halo}

Studies of the Galactic halo shed light on the formation of our
galaxy. The kinematics of the stellar halo are an important indicator of the
formation mechanisms involved. Some of the more recent studies that derive the
halo velocity ellipsoid in the inner halo include \citet{layden} who used
low-metallicity RR Lyraes, and \citet{gouldRR} who used low-metallicity
RR Lyraes as well as other low-metallicity stars. \citet{gouldRR} derive 
a halo rotation in prograde direction of
$34.3\pm8.7\,\kms$ with respect to the Galactic frame. This estimate, which assumes that
the Sun's velocity in the Galactic plane is $232\,\kms$ in the prograde
direction, applies to halo stars within 3 kpc.

{\it FAME} will be able to greatly improve the range, the resolution and the
accuracy of the halo velocity ellipsoid, thus allowing one to see any gradient 
in motion. {\it FAME} will be able to do this by observing many faint blue
horizontal branch stars in various directions of the sky. Horizontal branch
stars are especially favorable for mapping purposes as their nearly constant
luminosity permits a relatively precise estimate of their distances.

Candidate blue horizontal branch stars (BHB stars) can be effectively selected
from multi-band photometry. This was recently demonstrated by \citet{yanny}
using SDSS $u'g'r'$ photometry. Their study showed that A-colored stars
(A-stars) trace huge substructures in the halo. Unfortunately, other types of
blue stars, those with main sequence gravity (mostly field blue stragglers
(BSs)), have similar colors as BHB stars, and it is not clear whether
distinguishing between these two populations can be done based on photometry
alone. Mixing these two types of stars that have very different absolute
magnitudes (and with BS luminosities likely having a strong metallicity
dependence as well) makes it difficult to use them as distance indicators. One
certain way of distinguishing any individual star is by measuring the widths
of Balmer lines that indicate surface gravity (BHB stars have lower surface
gravity and narrower lines). However, spectroscopy will not be available in
the majority of cases, and we must consider other methods. The stars of the
two population with same apparent magnitudes lie at different distances, so
they will on average exhibit different proper motions. It is this feature that
we will use to distinguish between BS and BHB-star populations.

BHB stars with $R<15$ will already be included in the {\it FAME} input
catalog. In the 1/2 of the north polar cap for which SDSS data will be
available one can employ color criteria similar to those used by
\citet{yanny}. Selecting stars with $15<r^*<18$ (for blue stars SDSS $r^*$ is
quite close to $R$), and with dereddened colors
\be -0.3<g^*-r^*<0.1 \qquad
0.8<u^*-g^*<1.5 
\ee 
we derive a surface density of A-stars in the $\sim 500\,{\rm deg}^{2}$ of SDSS
Early Release Data (EDR) of $\sim 4.2\,{\rm deg}^{-2}$. This implies that
approximately 20,000 BHB star candidates will come from SDSS. For the rest of
the northern cap and the entire southern cap, one can try to retrieve BHB
candidates from some other all-sky catalog. We perform this exercise using USNO-A2.0, although the Guide Star Catalog 2 (GSC-2), which will have better calibrated photometry, will be better suited. Since in the case of USN0-A2.0 we have at
our disposal only two-band photometry which is quite crude (color error of
$\sim 0.4$ mag), we need to estimate the efficiency of getting A-star
candidates given some level of contamination from the many times more numerous
turn-off stars. To do this we matched SDSS EDR A-stars to USNO-A2.0 objects
and identified them on the USNO-A2.0 color-magnitude diagram (CMD) of all
stars within the SDSS EDR sky coverage. SDSS selected A-stars in this CMD lie
on the blue side, as expected, but are not clearly separated from the turn-off
star locus. Since USNO-A2.0 provides only single color on which to base the
selection, the blue portion of the USNO-A2.0 CMD also contains white dwarf
and quasar contaminants, as well as turn-off stars. In the case of SDSS selection these quasars and white dwarfs were eliminated using the
$u^*-g^*$ color. Although in the present context WDs and QSOs are considered
as contaminants, including them in the input catalog is useful for other
aspects of this project. As stated previously, QSOs can be distinguished by
their $K$-band excess, while WDs will stand out by their much higher proper
motions.

The locus of both the turn-off stars and the SDSS selected A-stars in the 
USNO-A2.0 CMD is tilted with a slope corresponding to the line,
\be B_{\rm USNO}-R_{\rm USNO} = a - 0.15 R_{\rm USNO},
\ee 
where $a$ is the zero point of the line. Now we can count the number of
$15<R_{\rm USNO}<18$ USNO-A2.0 stars left (blue-ward) of this line as we shift
this line red-ward, and at the same time noting how many USNO matches to SDSS
selected A-stars are included in the region to the left of the line. As we
move red-ward, both the number of real (SDSS selected) A-stars and the number
of USNO-A2.0 A-star candidates will increase. We find that the ratio of
USNO-A2.0 A-star candidates to SDSS A-stars remains constant for $a<2.4$, but
that the turn-off contamination increases rapidly further red-ward of this
point. For $a<2.4$ we find the ratio of USNO candidates to SDSS A-stars to be
12. Such selection retrieves $54\%$ of SDSS selected A-stars. To summarize, if
this selection is applied to 15,000 ${\rm deg}^{2}$ of the northern and southern
caps not covered by SDSS, one will end up with 340,000 candidates that will
contain 28,500 actual A-stars. Note again that many of these `contaminants'
will be QSOs and WDs which one would want anyway.

In the case that the {\it GALEX} UV sky survey becomes available, we would be able to
eliminate turn-off contaminants in USNO-A2.0 selected A-star
candidates. Alternatively, one can use {\it GALEX} to select A-star candidates
and then match them in USNO-A2.0 (or better yet GSC-2).

We now come to the question of distinguishing A-stars as either BHB stars or
BSs, using proper motions from {\it FAME}. For simplicity, we will assume that
all A-stars are located exactly in the direction of the Galactic pole, so that
only two of the three components of the motion are expressed. In one
direction, let us call it $y$, we will then see the rotational motion around
the center of the Galaxy. The velocity corresponding to this motion is
approximately $200\,\kms$, and the velocity dispersion in this direction
according to \citet{gouldRR} is $109\,\kms$. In the perpendicular direction
($x$), parallel to Sun--Galactic center vector, we will assume no bulk motion,
and the \citet{gouldRR} velocity dispersion of $160\,\kms$.

At each apparent magnitude both BHB stars and BSs are sampled, but their ratio,
even if constant in a given volume, is not independent of apparent
magnitude. If we assume, as for example implied by \citet{yanny}, that in a
given volume the field BSs outnumber field BHB stars 2:1, and that both 
populations fall off with Galactocentric distance as $r^{-3.5}$, then from our
vantage point 8 kpc from the center, at any given magnitude we will sample BSs
at a different Galactocentric distance than BHB stars, and the magnitude bin
will correspond to different volumes for the two types of stars because of
different heliocentric distances. Thus at $R=15$ we find $N_{\rm BS}/N_{\rm
BHB}= 0.17$, but at $R=19$ the ratio is 2.3. Other values can be deduced from
Table \ref{table:halo}.

Next we use Monte Carlo techniques to simulate observations of all A-stars that
one hopes to select using SDSS and USNO-A2.0, as previously described. We do
this for $R=15, 16, 17, 18,\, {\rm and}\, 19$. For each magnitude bin
($R\pm0.5$) we first estimate the number of BHB stars and BSs that {\it FAME}
will detect in that bin (given in Table \ref{table:halo}), then to BHB stars
we assign an absolute magnitude drawn randomly from $M_R=0.8\pm0.15$, while to
BSs we assign absolute magnitudes distributed as $M_R=3.2\pm0.5$, where the
dispersion is meant to reflect the range of absolute magnitudes at a given
color. This will determine the distance to that star. Then each star is
assigned two components of physical velocity drawn from the halo velocity
distribution described earlier in this section. Using the distance, we convert
this velocity into a true proper motion. The observed proper motion is then
obtained by adding in a measurement error based on the {\it FAME} accuracy
$\sigma_\mu$ from \S\ 2.1.

Since at any given magnitude BSs lie three times closer than BHB stars, in each
apparent magnitude bin there will be two peaks in proper motion, corresponding
to BSs and BHB stars. Because of proximity and because of a greater spread of
absolute magnitudes, the distribution of proper motions corresponding to BSs
will be wider at a given magnitude than that of BHB stars. It is the position
of the peak of BHB stars proper motion distribution in $y$-direction that will
yield the halo rotational velocity. To find out how well this peak can be
determined in the face of BS contamination, we calculate errors in fitting
2-dimensional Gaussians to each of the two peaks. Each Gaussian is defined by
six parameters -- two for the $x$ and $y$ center of the peak, three for two
diagonal ($\sigma_{xx}, \sigma_{yy}$), and one off-diagonal ($\sigma_{xy}$)
term in the covariance matrix describing peak widths, and one corresponding to
the number of stars (amplitude). We calculate the errors of these parameters
from our simulated measurements.

\begin{figure}
\plotone{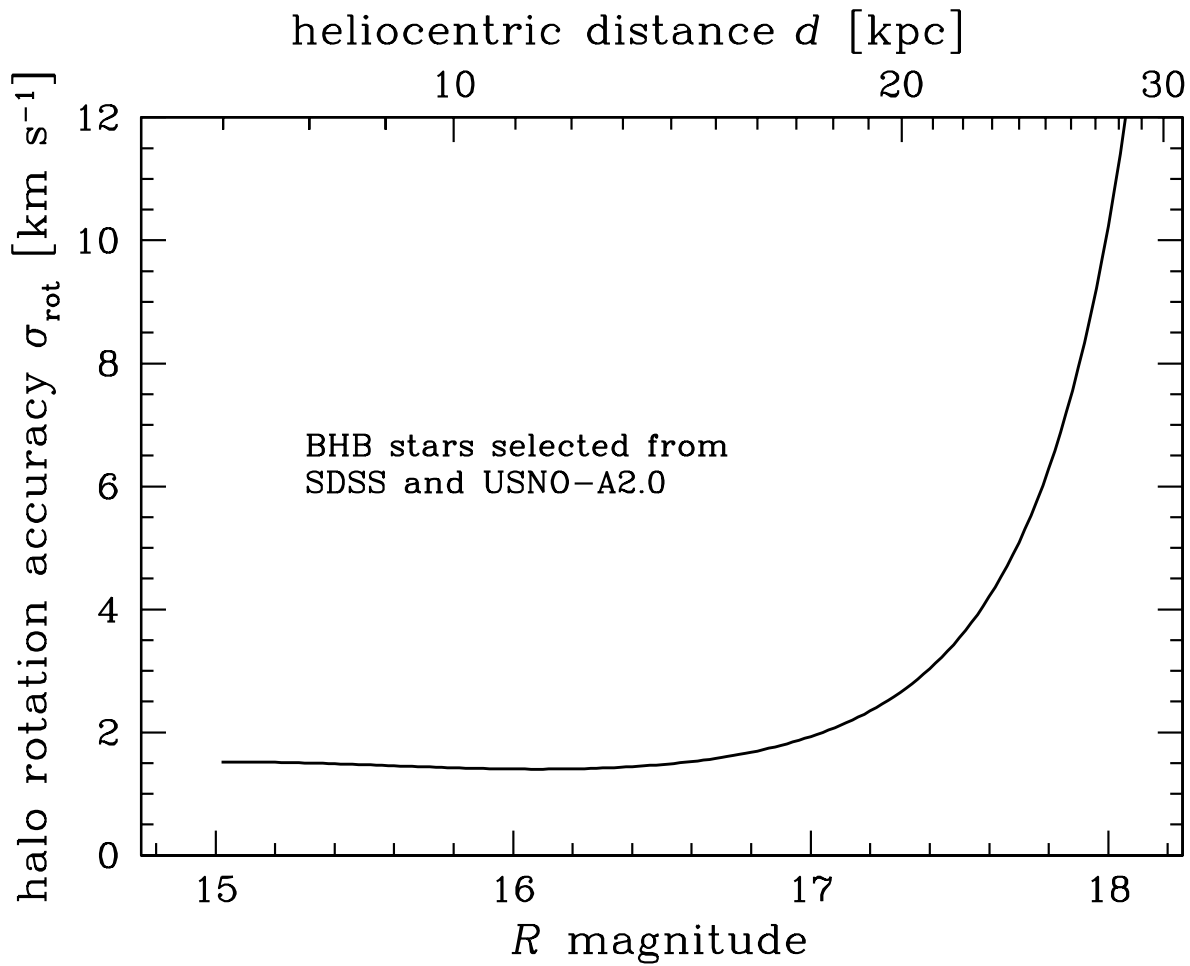}
\caption{The kinematics of the halo, and especially its rotation, can be measured by selecting field blue horizontal branch stars, which have good photometric distances, and measuring their mean proper motion. We show the precision of this estimate, assuming {\it FAME} sensitivities, as a function of BHB stars' $R$ magnitude, or equivalently, their heliocentric distance. The sample is binned in 1-mag bins. Deterioration beyond $R\sim18$ occurs mainly because the color-selected sample of A-stars becomes dominated by blue stragglers. \label{fig:halo}}
\end{figure}

The results are summarized in Table \ref{table:halo}. The second column lists
the distance of BHB stars of magnitude $R$, which is given in the first
column. The third and fourth columns list the number of BHB stars and BSs that
we expect {\it FAME} to measure in the two caps ($|b|>30\degr$). The next two
columns list the errors that we derive for the position of the BHB proper
motion peak, and columns 7 and 8 list estimates of the uncertainty of the BHB
stars velocity dispersion along $y$ and $x$ directions. The final column,
derived from columns 2 and 5, shows the expected error in the stellar halo
rotation velocity. In Figure \ref{fig:halo} we present $\sigma_{\rm rot}$ as a
function of $R$ or $d$. One can see that within 20 kpc {\it FAME} achieves a
precision of halo rotation measurement of $\lesssim 2\,\kms$ if the data are binned in
1-mag steps. Better spatial sampling (resolution) can be achieved by choosing
smaller bins, but with correspondingly larger errors. At distances to $\sim
30\,\kpc$ quite good estimates of stellar halo rotation can still be
made. Only farther out do the measurement errors and the preponderance of BSs
of same magnitude as the BHB stars preclude obtaining a useful result. Note that here we assume that the BHB stars' luminosity will be very well determined locally by {{\it FAME}'s} trigonometric parallaxes, and that this luminosity does not depend on distance from the plane. To get motions relative to the 
Galactic frame, it will of course be necessary to subtract the Sun's 
circular velocity, which will have been determined as described in \S 4.1.

A more sophisticated analysis would show that the halo rotation can be mapped
in two dimensions, instead of one as outlined here. Such probing of the halo
potential would place more constraints on the dark halo models. Additional
information on the still debated shape of the dark matter halo (flatness [Olling \& Merrifield 1998, 2001] and
triaxiality) will be gleaned from the three-dimensional distribution of stars
determined to be BHB stars from {\it FAME} proper motions. An effect that might introduce systematic error in distances and thus the velocities is if BHB 
stars change in luminosity (due to age and metallicity effects) as we move above the plane. This problem might be eliminated by imposing an axial symmetry--we would require $v_{\rm rot}$ to be constant at a given Galactocentric radius, regardless of the direction. Such a treatment requires an analysis 
that is beyond the scope of this paper. 

In the above analysis we have used BHB stars to probe halo kinematics and
considered BSs as a contaminating factor. We note that with the halo motion
determined by BHB stars, the measured proper motions of BSs can be used to
derive their luminosity calibration and distribution of metallicities.

\subsection{Substructure in the Galactic Halo}

Recently discovered clumps of RR Lyrae stars \citep{ivezic}, and A-colored
stars \citep{yanny} in the halo indicate that the Galaxy formation mechanism
might be more complex than previously envisaged, and that accretion or merging
of small galaxies might have played a crucial role. Some models even suggest
that all halo stars come from disrupted satellite galaxies. The structures
found by \citet{yanny} lie at Galactocentric distances of 30 to 50 kpc, and
extend over many kiloparsecs. The structures are seen as overdensities of BHB
stars and BSs, i.e., only positional information is used. The third coordinate,
the distance, is compromised because of difficulties with BHB/BS distinction,
leading to smearing of the features. Besides being clumped in space, the stars
originating from the same disrupted satellite should cluster in velocity
space as well. In fact, the velocity information is conserved much better than
the spatial information, and it is possible to associate stars that are widely
separated on the sky and that have mixed spatially with other
streams. \citet{helmi} have shown that 10 Gyr after a merging event the
spatial distribution of stars in the halo will be very smooth, while hundreds
of halo streams, strongly ($\sigma_v<5\,\kms$) clustered in velocity space
will still be present.

{\it FAME} observations will yield two components of velocity that can be used
to detect substructure in the halo. The old halo streams predicted by
\citet{helmi} will require local samples of subdwarfs. The best sensitivity
for detecting such streams comes from analyzing the proper motions of
G-subdwarfs (halo turn-off stars) which are bright and numerous. \citet{helmi}
suggest that the velocity accuracy needed to resolve individual streams is
$<5\,\kms$, but that they will be detectable at several times that
accuracy. We believe that such studies will be possible with stars selected as
part of the {\it FAME} main survey ($R<15$). Including fainter stars might not
be useful as the velocity accuracy would be limited by inaccurate distances.
This is indicated by \citet{helmi2} who find that {\it FAME} will be able to
distinguish 15\% of the nearby halo streams. Their sample of halo giants is
limited by parallax errors to $V<12.5$, and the velocity measurements are
augmented with an assumed ground-based radial velocity survey.

Bigger structures (tidal streams and remnants of recently disrupted
satellites) might be detectable in the $R>15$ BHB star sample discussed
previously. The specific structures found by \citet{yanny} may be too distant
to be detected by {\it FAME}, yet their survey covers only $1\%$ of sky, so
more nearby clumps are likely to be present in the rest of the sky. To find
them one should aim at $\sim 20 \,\kms$ accuracy per star. With the good
photometric parallaxes achievable for BHB stars, substructure mapping might be
possible to distances of 10 kpc (corresponding to $R_{\rm BHB}\approx16$).

\subsection{Proper Motion of LMC, SMC, Dwarf Spheroidals and Globular Clusters}

Proper motions of the Magellanic Clouds, dwarf spheroidal galaxies and globular clusters
add to our knowledge of the kinematics of these objects and the extent and
properties of Milky Way's dark matter halo (e.g \citealt{wilk,koch}). 
{\it FAME} will in some cases
observe proper motions of bright ($R<15$) stars in these systems, thus
allowing the proper motion of the entire system to be determined. For the LMC and SMC, at least, the 
accuracy will only be limited by the accuracy of the reference
frame. Hence, a precise determination of reference frame rotation using
quasars, as described in \S 3, will again be of importance. Let us illustrate
this with the case of the LMC. \citet{gouldLMC} has shown that a kinematic
distance to the LMC can be obtained independent of any distance calibrators
such as RR Lyraes or Cepheids. The method requires a measurement of the proper
motion of the LMC. Using {\it FAME} observations of $\sim 8600$ $13<V<15$ LMC
stars, \citet{gouldLMC} estimates that the proper motion of the LMC can be
determined to $2\,\uasyr$, or $0.2\%$ accuracy. However, this unprecedented
accuracy is limited by frame accuracy, and would be compromised if only $R<15$
quasars were used to define the frame.
 
\section{Faint Nearby Stars}

\subsection{Late-M and L Dwarfs}

In recent years there has been a major breakthrough in the discovery and study
of late-type stars and substellar objects (brown dwarfs). Wide-area sky
surveys, that either go much fainter than the previous ones (SDSS), or image
the sky in the near-infrared where these cool stars emit most of their energy
(2MASS, DENIS), allow for the first time a great number of these objects to be
discovered. These discoveries led to the introduction of two new spectral
classes: L dwarfs that are cooler than the latest M dwarfs and some of which
might have substellar mass, and T dwarfs that exhibit methane absorption 
and which are certainly
substellar. Currently, over 100 L dwarfs have been found, and some two dozen T
dwarfs \citep{burg,leggett,kirk}.

Accurate distances are a key factor in understanding the structure and
evolution of these objects. Besides establishing luminosity calibration,
distances would allow better constraints on other key parameters like the
radius and the temperature. Distances are currently available for about
20 L dwarfs, and a few T dwarfs. In some cases a precise distance is
known only because the object is a companion to a star with a
measured trigonometric parallax. Distances will also help constrain the
luminosity function of these objects.

Even for the more easily discovered late M dwarfs, the faint end of the main
sequence ($V-I>3$) is defined with the trigonometric parallaxes of only some
30 stars, mostly coming from the USNO CCD parallax program \citep{monet}. This is
too few to accurately describe the differences in absolute magnitudes of
different populations of stars.

One of the already recognized mission goals of {\it FAME} is to refine the
absolute magnitude calibration in the entire HR diagram, including the faint
main sequence stars. However, selecting only $R<15$ stars will leave
unmeasured many stars for which good ($<10\%$ accuracy) parallaxes are within
the reach of {\it FAME}, with the ratio of unobserved to observed stars
getting worse as one goes further down the main sequence.

\begin{figure}
\plotone{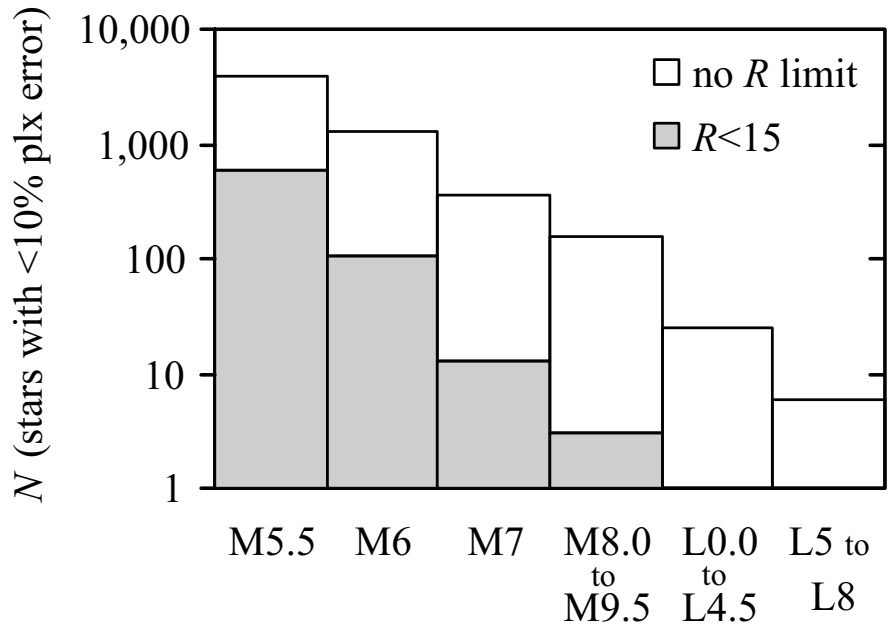}
\caption{The histogram shows the number of late-type dwarfs of different spectral classes for which one can measure parallaxes with fractional error $<10\%$,
assuming {\it FAME} sensitivities. If objects are selected beyond the main survey magnitude limit of $R=15$ (blank vs.\ grey bars), there is a significant increase in the number of good parallaxes. \label{fig:MLT}}
\end{figure}

In Figure \ref{fig:MLT} we compare the number of stars for which {\it FAME}
will be able to measure parallaxes with $<10\%$ errors if no magnitude limit
is imposed, compared to the number of such stars brighter than $R=15$. Our
first bin is dM5.5 stars (equivalent to $M_V=14$). There are intrinsically
brighter stars (late K and early M dwarfs) for which including $R>15$ stars
would also produce many additional good parallaxes. However, the number of
such stars with $R<15$ is already high enough to allow calibration at the
level of their intrinsic scatter. In calculating the number of stars in Figure
\ref{fig:MLT} we use current estimates from the literature for the colors,
magnitudes, and space densities for late-M, L and T dwarfs, coupled with {\it
FAME} astrometric precision. The numbers for L dwarfs are less
certain (by a factor of two) because of their poorly known number
densities. Since the stars fainter than $R=15$ must be deliberately selected
to be included as nearby star candidates, and this selection might be
difficult close to the Galactic plane, our estimate is based on the assumption
that M dwarfs will be selected only if their Galactic latitude satisfies
$|b|>20\degr$. On the other hand, we assumed that L dwarf candidates
will be followed up and confirmed in all regions of the sky.

For dM5.5 stars, discarding the $R=15$ magnitude limit augments the number of
parallaxes from 600 to 4000 by extending the volume probed from 30 to 64
pc. For the very latest M-dwarfs (M7 to M9.5) one will obtain 500 measurements, compared to only 16 for $R<15$. Finally about 5 L dwarfs are present in the $R>15$, $\sigma_\pi/\pi<0.1$ sample, whereas the $R<15$ limit excludes all L dwarfs. It is obvious that, except for L dwarfs, which are very faint and
whose parallaxes are possibly better determined by other means (ground-based
campaigns, space telescopes), accepting this magnitude-limit-free selection
leads to great improvements in calibrating low-mass stars. We also find that T dwarfs are outside of reach of {\it FAME}, although a few lower precision measurements could possibly be made.

While most of the nearby L and T dwarfs will be found and confirmed by
dedicated searches, late M dwarfs need to be selected from some
catalog. Fortunately, 2MASS will contain all the late M dwarfs that {\it FAME}
can observe. We estimate the number of $M_V>14$ candidates by running a search
through the 2nd Incremental Release of 2MASS PSC. We use the following search
criteria: $H-K>0.3$, $J-H>0.3$ and $R_{\rm USNO}-K_s>3.5$ (or no USNO
detection) to select late M dwarfs based on color; $J-K_s \leqslant 4/3(H-K_s)
+ 0.25$ to eliminate M giants \citep{gizis}; $J\leqslant14.5$ to ensure
selection of nearby objects, and $R_{\rm USNO}>15$ to exclude stars already in
the {\it FAME} main survey list. The search is further restricted to 
$|b|>20\degr$ to
exclude contaminants in the Galactic plane. We find the average density of
candidates to be $2.3\,{\rm deg}^{-2}$, implying 64,000 total candidates,
which is a modest addition to the {\it FAME} input catalog. In 1/5 of this
area SDSS photometry will also be available which will permit refinement of
the selection The selection can be further augmented with GSC-2, which will 
contain photometry similar to $I$ band. Any contaminants in this sample will 
at the end be obvious from
the {\it FAME} parallax measurements themselves.

\subsection{White Dwarfs}

White dwarfs are one of the most numerous stellar populations in our
neighborhood. Therefore they represent obvious targets for getting precise
distances. A large sample of parallaxes would allow one to study the white
dwarf mass function. The mass of a white dwarf is related to the mass of the
progenitor and can thus serve as an indicator of the IMF at different
epochs. Current measurements of the mass function are still somewhat
controversial both in terms of the peak mass (estimates range from
$0.48M_\sun$ to $0.72M_\sun$) and also concerning the width of the mass
function (for recent results see Silvestri et el.\ 2001). At a given
temperature (or color), WDs exhibit a range of absolute magnitudes reflecting
a range of radii, since the absolute magnitude is then primarily a function of
the white dwarf's radius. Then the mass-radius relation, which is very well
determined theoretically, allows one to find the masses. Other methods of
getting masses either rely on a few binary systems or are of low precision
(gravitational redshift). Obtaining luminosities of the coolest white dwarfs
is also useful for testing the predicted cooling scenarios.

\begin{figure}
\plotone{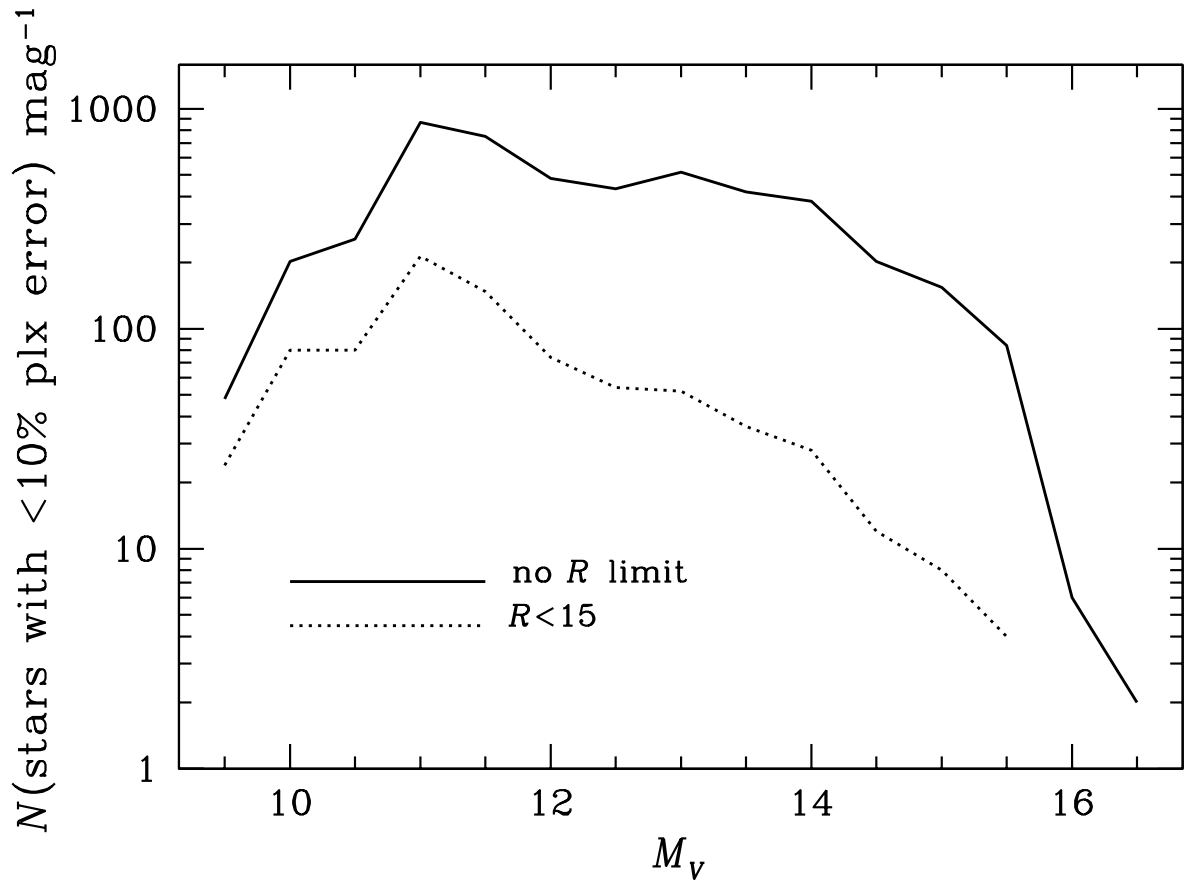}
\caption{Observing WDs fainter than $R=15$ greatly increases the number of good ($<10\%$ fractional accuracy) parallaxes. The number (per magnitude) of good parallaxes (assuming {\it FAME} sensitivities) in the two cases is given as a function of WD absolute magnitude. \label{fig:wd}}
\end{figure}

Currently $\sim200$ precise (accuracy better than 10\%) trigonometric
 parallaxes of white dwarfs are known \citep{mccook}. Using the white dwarf
 luminosity function from \citet{ldm} and our estimates of {\it FAME} parallax
 precision, we estimate that the {\it FAME} survey of $R<15$ stars will
 produce $<10\%$ parallaxes of 400 white dwarfs. However, as in the case of
 late-M and L dwarfs, one can increase this number by including in the input
 catalog objects fainter than $R=15$. Then the total number of white dwarfs
 with precise parallaxes measured by {\it FAME} rises to 2400, i.e., a
 six-fold increase.  This increase is
 accentuated for fainter white dwarfs, as can be seen in Figure \ref{fig:wd}. For example, for $M_V>13$ ($T\lesssim
 8500\,{\rm K}$) the increase with respect to $R<15$ sample is
 thirteen-fold. Such numbers would allow mass functions to be constructed as a
 function of age, and for different atmospheric compositions.

The magnitude limit for 10\% parallaxes will range from $R=15.5$ for the
bluest white dwarfs ($M_V=9.5$), to $R=17.6$ for the reddest
($M_V=16.5$). These faint WDs must first be selected and then added to the
input catalog. Currently known white dwarfs account for $\lesssim40\%$ of the
expected $R>15$ sample, based on white dwarfs with photometric information
from \citet{mccook}. The most efficient way of selecting white dwarfs is by
using an all-sky proper motion catalog. Then, the white dwarfs are selected
from a reduced proper-motion diagram in which the reduced proper motion, $H_m =
m + 5\log\mu +5$, is plotted against color. For a stellar population with
approximately the same transverse velocities, the proper motion becomes a proxy
for the distance, and the reduced proper motion a proxy for the absolute
magnitude. Because the white dwarfs are separated by $\sim10$ mag from the main
sequence, they will stand out in a reduced proper-motion diagram by the same
amount, since the two populations have on average the same transverse
velocities. For redder white dwarfs, there will be some contamination from
subdwarfs, since due to their greater velocity these stars will move down into
the white dwarf region. However, even if some subdwarfs `contaminate' the
white dwarf observing list, these stars will be of considerable interest in
their own right.  The only currently available all-sky proper motion catalog
that contains faint stars, the \citet{nltt} NLTT catalog, 
has a proper motion cutoff that
is too high to include most of the white dwarfs with typical disk velocities
that lie within the volume accessible to {\it FAME}. Thus, it will be necessary
to draw white dwarf candidates from the yet to be released GSC-2 or USNO-B
that will include all detectable proper motions with typical accuracy of
$5\,\masyr$. This will be precise enough to select even the farthest WDs that
produce 10\% parallax measurement, since they typically move at
$40\,\masyr$. In the part of the sky observed by SDSS, hot ($B-V<0.3$) white
dwarf candidates can be selected solely based on their very blue color in {\it
all} SDSS bands (as distinct from other blue objects like the quasars and
A-type stars, see Fan 1999), while SDSS photometry of red candidates from
GSC-2/USNO-B will be good enough to distinguish subdwarfs from WDs.
GSC-2/USNO-B white dwarf candidates can additionally be cross-correlated with
{\it GALEX} UV sources, assuming that the {\it GALEX} survey is complete by
the time required to define the input catalog. Some nearby white dwarfs might
also end up in the input catalog as `contaminants' from other projects
suggested here (when selecting quasars and blue horizontal branch
stars). Generally, we do not expect the white dwarf candidates to represent a
significant increase in the input catalog size. It will be useful for
spectroscopic follow-up to commence as soon as candidate nearby white dwarfs
are identified. Besides offering spectroscopic confirmation, spectroscopy is
necessary to establish white dwarf atmosphere composition that in turn allows
parallax data to be interpreted correctly.

\subsection{Planetary and Brown Dwarf Companions of M Dwarfs}

	Here we examine how adding faint $(R>15)$ M dwarfs to the
{\it FAME} input catalog can affect its sensitivity to brown dwarfs 
and planets.  For completeness, we show {\it FAME}'s 
sensitivity to companions of earlier type dwarfs as well.  To
perform this calculation we adopt the vertical disk profile
of \citet{zheng} (Table 3, all data, CMR-2, sech$^2$ model),
and adopt their (CMR-2) M dwarf luminosity function (LF) as well.
For earlier type stars we use the LF of \citet{bessell}.
Note that the
\citet{zheng} LF is uncorrected for binaries which is appropriate
because most M dwarfs that are companions of brighter primaries will
not be individually resolved by {\it FAME}.

\begin{figure}
\plotone{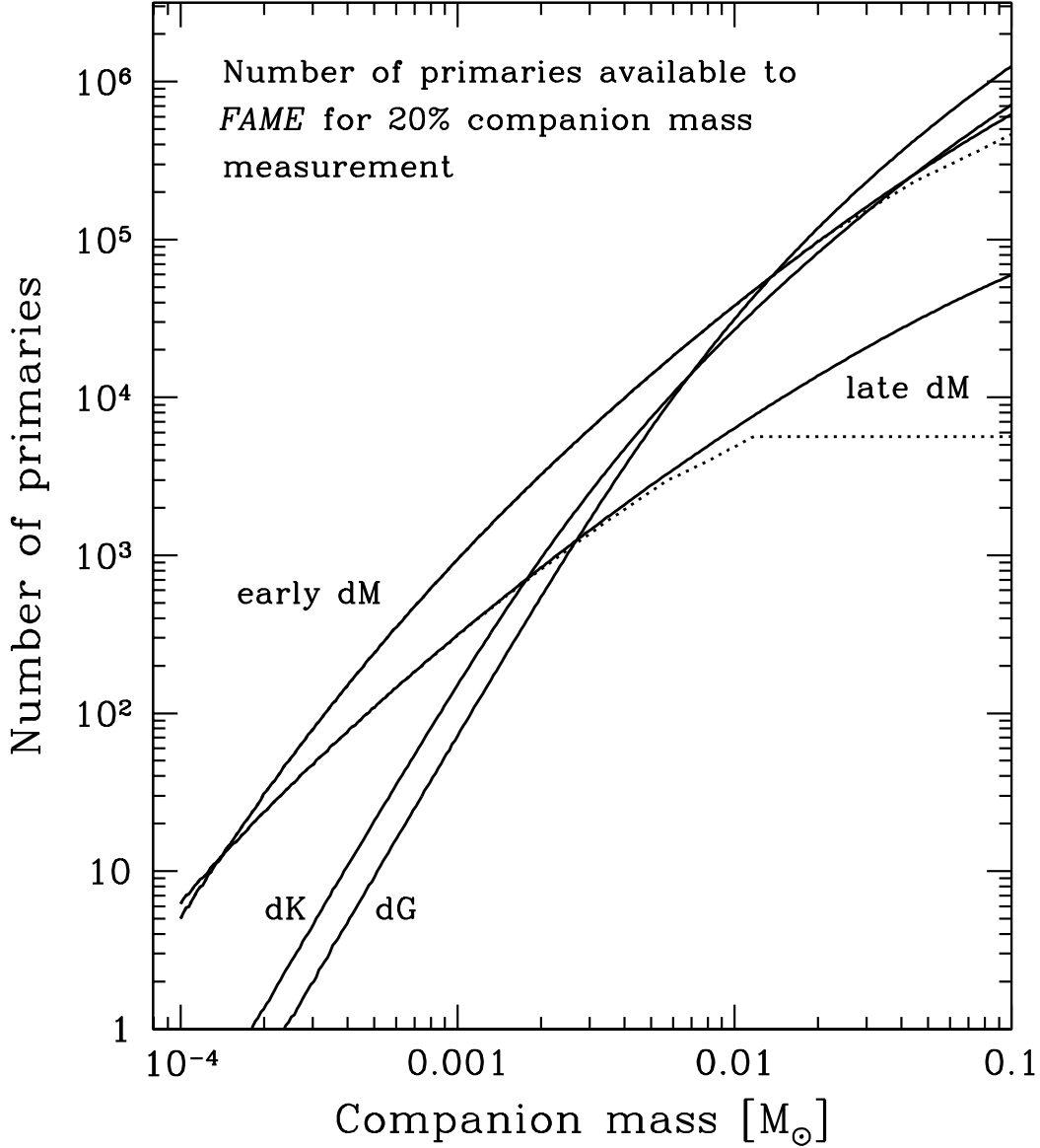}
\caption{{\it FAME} sensitivity to planetary and brown dwarf companions.
The solid curves show the number of stars for which {\it FAME}
could detect companions in $P=5\,$yr orbits as a function of 
companion mass.  The curves separately show 
G dwarfs ($3.5<M_V<5.5$)
K dwarfs ($5.5<M_V<7.5$)
early M dwarfs ($7.5<M_V<12.5$), and
late M dwarfs ($12.5<M_V<18.5$).  These curves assume that 
M dwarfs with $R<18$ will be included in the input catalog.  The
effect of excluding stars with $R>15$ is shown by the dashed curves.
For shorter periods, $P$, the curves should be displaced to the right by
$(2/3)\log(5\,{\rm yr}/P)$. \label{fig:bds}}
\end{figure}

	Figure \ref{fig:bds} shows the number of stars whose companions of
a given mass would be detectable by {\it FAME}.  It specifically
assumes a $20\%$ mass measurement threshold, 
and an orbital period
of $P=5\,$yr.  The effect of requiring $10\,\sigma$ detections
can be approximately gauged by displacing all the curves by
0.3 dex to the right.  For shorter periods, one should displace
the curves to the right by $(2/3)\log(5\,{\rm yr}/P)$.  For
example, for $P=2.5\,$yr, the curves should be displaced by 0.2 dex.
Figure  \ref{fig:bds} illustrates that the major effect of including faint
M dwarfs is to enhance sensitivity to brown dwarf
companions of late M dwarfs.   For companions just below the
hydrogen burning limit, there is an order of magnitude
improvement.  Note that {\it FAME} will be sensitive to Jupiter
mass companions of 1300 M stars and to 10-Jupiter mass companions
of 45,000 M stars, and of an even larger number of G and K stars.  
However, these remarkable sensitivities
will not be significantly improved by inclusion of faint stars.

\subsection{Miscellaneous Stars}

Our targeted search to include late M dwarfs and white dwarfs should ensure
that most such nearby objects are included, however there is a possibility
that some will be missed. Therefore, and also for the sake of ensuring a
volume-complete sample of the nearest stars (within 25 pc), one should make sure that all objects from the Catalogue of Nearby Stars
\citep[CNS3]{cns3} are in the input catalog. 
Furthermore, it will be useful to include the complete NLTT. Besides still
representing a reservoir of new nearby stars \citep{scholz}, NLTT contains
many stars with intrinsically high space velocities -- subdwarfs, and
some halo white dwarfs. The luminosity calibration and true space velocity of
these objects is of great interest in many Galaxy evolution studies. 
Finally, a few other interesting objects found in other surveys (again, like
high-velocity white dwarfs) should be included. All of these targets will
comprise a negligible fraction of the {\it FAME} input catalog. In fact, there are relatively few specially selected and catalogued stars (of any magnitude), and all of them will most likely end up in the {\it FAME} input catalog (D. Monet 2001, private communication).

\section{Impact on Main Survey}

Depending on which sky surveys will be available several years from now, and
on exactly which of the various selection strategies that we have elaborated
were adopted, the faint samples discussed above would total of order
$\mathcal{O}(10^6)$ objects, or about 2.5\% of the {\it FAME} catalog.
Since {\it FAME} is fundamentally limited by data transmission rate, these
catalog entries would come at the expense of the magnitude-limited sample,
e.g., by reducing the magnitude limit by 0.03 mag.  This would not 
significantly affect any of {\it FAME}'s primary science goals.

\section{{\it DIVA} and {\it GAIA}}

As discussed in \S\ 1.2 and \S\ 2.1, the characteristics of {\it DIVA} are 
close
enough to those of {\it FAME} that the methods and arguments presented here can
be applied directly to {\it DIVA}.  Indeed, if the constants $c_{\rm RN}$
in equations (\ref{eqn:cons}) and (\ref{eqn:consdiva}) were the same,
the entire analysis of {\it FAME} could be applied directly to {\it DIVA}
simply by rescaling the errors (by a factor 4.0 for parallaxes and 6.4
for proper motions).  Since the read-noise limit sets in 0.64 mag brighter
for {\it DIVA} than {\it FAME}, pushing beyond the magnitude limit yields
slightly less additional science.  Nevertheless,
qualitatively the effect is the same.

While the methods presented here can ultimately be used to choose faint
targets for {\it GAIA}, it is not practical to do so at the present time,
primarily because the luminosity functions of the relevant target populations
are not sufficiently well understood.  For example, one could in general
advocate observing quasars fainter than $V=20$ {\rm GAIA} limit in order
to improve the reference frame.  However, the net frame error from 
observing a flux-limited sample of quasars scales as
\be
\label{eqn:qsolim}
\sigma_{\rm frame}^{-2} \propto
\int_{F_{\rm lim}}^\infty d F\,[\sigma(F)]^{-2}\Phi(F),
\ee
where $\Phi(F)$ is the quasar LF as a function of flux.
Figure \ref{fig:qso} shows that there is a substantial improvement in
the {\it FAME} frame when going from $R=18$ to $R=19$, even though
the readnoise limit applies, $\sigma(F)\propto F^{-1}$.  This is because
the LF is very steep, $\Phi\propto F^{-2.8}$, so the integrand in
equation (\ref{eqn:qsolim}) scales as $\propto F^{-0.8}dF$.  The
LF is believed to flatten considerably (index change $\sim 1$) at $R\sim 19$
\citep{hs} which, if true, would virtually eliminate the value of going
to still fainter quasars.  However, this break in the LF is based largely
on photographic data and occurs just at the magnitude at which the quasars fall
below the sky.  It therefore may be subject to revision when wide-area
CCD data come in from SDSS.  On the other hand, if {\it GAIA} ultimately 
manages to operate in the photon-limited (rather than currently foreseen
readnoise-limited) regime, this would approximately compensate for
the break in the slope.  Hence there is simply too much uncertainty
in both the sample characteristics and the mission characteristics
to make a reliable judgement on the usefulness of this type of extension.
Similar remarks apply to pushing {\it GAIA} beyond its magnitude limit
in pursuit of L and T dwarfs, halo white dwarfs, and indeed any class of
objects that one might contemplate.

It should be noted that unlike {\it FAME}, {\it DIVA} and {\it GAIA} do not
use input catalogs, but cut off faint stars using an on-board
photometer. Nevertheless, in those cases, like in the case of {\it FAME}, a
list of faint objects that should not be cut off should be produced in advance
and provided to a satellite.

\section{Conclusions}

{\it Hipparcos} added faint stars to its observing list that were 
approximately equal in number to the catalog's magnitude-limited 
component, thereby permitting a wide variety of scientific investigations
that would not otherwise have been possible.  Future astrometric missions
can also dramatically augment their science capability by adding targets
beyond their magnitude limit.  We have specifically analyzed the 
potential of {\it FAME} in this regard, and found that the addition of
$\mathcal{O}(10^6)$ targets (2.5\% of the total) can improve the
precision of the reference frame by a factor $\sim 2.5$, yield precise
($2\,\kms$) measurements of halo rotation out to 25 kpc, greatly improve
the proper-motion measurements of Galactic satellites, and
increase the samples of L dwarfs, M dwarfs, and white dwarfs 
with good parallaxes by an order of magnitude.  We have shown that
assembling the various target lists required to make these observations
is not trivial, but have presented viable approaches to selection in
each case.

The methods that we have presented could easily be applied to {\it DIVA}.
Application to {\it GAIA}
should be deferred until closer to launch when both the mission characteristics
and the potential target characteristics are better understood.

\acknowledgments This publication makes use of catalogs from the Astronomical
Data Center at NASA Goddard Space Flight Center, VizieR Catalogue Service in
Strasbourg, and data products from the Two Micron All Sky Survey, which is a
joint project of the University of Massachusetts and the Infrared Processing
and Analysis Center/California Institute of Technology, funded by the NASA and
the NSF. It also uses services of SDSS Archive, for which funding for the
creation and distribution has been provided by the Alfred P. Sloan Foundation,
the Participating Institutions, the National Aeronautics and Space
Administration, the National Science Foundation, the U.S. Department of
Energy, the Japanese Monbukagakusho, and the Max Planck Society. The SDSS Web
site is \url{http://www.sdss.org/}. SDSS is a joint project of The University of
Chicago, Fermilab, the Institute for Advanced Study, the Japan Participation
Group, The Johns Hopkins University, the Max-Planck-Institute for Astronomy
(MPIA), the Max-Planck-Institute for Astrophysics (MPA), New Mexico State
University, Princeton University, the United States Naval Observatory, and the
University of Washington. This work was supported in part by Jet Propulsion
Laboratory (JPL) contract 1226901.

\clearpage

\clearpage

\clearpage

\clearpage

\clearpage

\clearpage
\begin{deluxetable}{l r r r r r r r r}
\tabletypesize{\footnotesize}
\tablecaption{Rotation of the halo from BHB stars. \label{table:halo}}
\tablewidth{0pt}
\tablehead{
\colhead{$R$} & 
\colhead{$d$}   & 
\colhead{$N_{\rm BHB}$}   &
\colhead{$N_{\rm BS}$} &
\colhead{$\sigma(\mu_y)$}  & 
\colhead{$\sigma(\mu_x)$} & 
\colhead{$\sigma_{yy}$} &
\colhead{$\sigma_{xx}$}     & 
\colhead{$\sigma_{\rm rot}$} \\
\colhead{mag} &
\colhead{kpc} &
\colhead{$10^3$} &
\colhead{$10^3$} &
\colhead{$\uasyr$} &
\colhead{$\uasyr$} &
\colhead{$\uasyr$} &
\colhead{$\uasyr$} &
\colhead{$\kms$} 
}
\startdata
15 &  6.9 & 6.3 &  1.1 &   46 &   66 &   250 &   502   &  1.52 \\
16 & 11.0 & 8.4 &  2.8 &   27 &   37 &    98 &   188   &  1.41 \\
17 & 17.4 & 9.0 &  6.7 &   24 &   28 &    68 &   107   &  1.94 \\
18 & 27.5 & 7.9 & 11.6 &   79 &   67 &   239 &   321   & 10.29 \\
19 & 43.7 & 9.3 & 21.6 & 2632 & 1784 &  6601 & \nodata & 545   \\
\enddata
\end{deluxetable}


\clearpage

\begin{thebibliography}{}
\bibitem[Backer \& Sramek(1999)]{backer} Backer, D.~C.~\& 
Sramek, R.~A.\ 1999, \apj, 524, 805 
\bibitem[Barkhouse \& Hall(2001)]{bark} Barkhouse, W.~A.~\& 
Hall, P.~B.\ 2001, \aj, 121, 2843 
\bibitem[Bessell \& Stringfellow(1993)]{bessell} Bessell, 
M.~S.~\& Stringfellow, G.~S.\ 1993, \araa, 31, 433 
\bibitem[Binney(1995)]{binney} Binney, J.\ 1995, IAU 
Symp.~166: Astronomical and Astrophysical Objectives of Sub-Milliarcsecond 
Optical Astrometry, 166, 239 
\bibitem[Boyle et al.(2000)]{2qz} Boyle, B.~J., Shanks, T., 
Croom, S.~M., Smith, R.~J., Miller, L., Loaring, N., \& Heymans, C.\ 2000, 
\mnras, 317, 1014 
\bibitem[Burgasser et al.(2002)]{burg} Burgasser, A.~J.~et 
al.\ 2002, \apj, 564, 421 
\bibitem[ESA(1997)]{hip} European Space Agency (ESA). 1997, The Hipparcos and Tycho Catalogues (SP-1200; Noordwijk: ESA)
\bibitem[Fan(1999)]{fan} Fan, X.\ 1999, \aj, 117, 2528 
\bibitem[Gizis et al.(2000)]{gizis} Gizis, J.~E., Monet, 
D.~G., Reid, I.~N., Kirkpatrick, J.~D., Liebert, J., \& Williams, R.~J.\ 
2000, \aj, 120, 1085 
\bibitem[Gliese \& Jahreiss(1991)]{cns3} Gliese, W., \& Jahreiss, H.\ 
1991, Preliminary Version of the Third Catalogue of Nearby Stars (Astron. Rechen-Institut, Heidelberg)
\bibitem[Gould(2000)]{gouldLMC} Gould, A.\ 2000, \apj, 528, 156 
\bibitem[Gould \& Popowski(1998)]{gouldRR} Gould, A.~\& 
Popowski, P.\ 1998, \apj, 508, 844 
\bibitem[Hartwick \& Schade(1990)]{hs} Hartwick, 
F.~D.~A.~\& Schade, D.\ 1990, \araa, 28, 437
\bibitem[Helmi \& de Zeeuw(2000)]{helmi2} Helmi, A.~\& 
de Zeeuw, P.\ 2000, \mnras, 319, 657 
\bibitem[Helmi \& White(1999)]{helmi} Helmi, A.~\& White, 
S.~D.~M.\ 1999, \mnras, 307, 495 
\bibitem[Ivezi{\' c} et al.(2000)]{ivezic} Ivezi{\' c}, {\v 
Z}.~et al.\ 2000, \aj, 120, 963
\bibitem[Kirkpatrick et al.(2000)]{kirk} Kirkpatrick, 
J.~D.~et al.\ 2000, \aj, 120, 447
\bibitem[Kochanek(1996)]{koch} Kochanek, C.~S.\ 1996, \apj, 
457, 228 
\bibitem[Layden et al.(1996)]{layden} Layden, A.~C., Hanson, 
R.~B., Hawley, S.~L., Klemola, A.~R., \& Hanley, C.~J.\ 1996, \aj, 112, 2110 
\bibitem[Leggett et al.(2002)]{leggett} Leggett, S.~K.~et al.\ 
2002, \apj, 564, 452 
\bibitem[Liebert, Dahn, \& Monet(1988)]{ldm} Liebert, J., 
Dahn, C.~C., \& Monet, D.~G.\ 1988, \apj, 332, 891 
\bibitem[Luyten (1979, 1980)]{nltt} Luyten, W.~J.\ 1979, 1980, New Luyten Catalogue of Stars with Proper Motions Larger than Two Tenths of an Arcsecond (Minneapolis: University of Minnesota Press) 
\bibitem[McCook \& Sion(1999)]{mccook} McCook, G.~P.~\& Sion, 
E.~M.\ 1999, \apjs, 121, 1 
\bibitem[Monet et al.(1992)]{monet} Monet, D.~G., Dahn, 
C.~C., Vrba, F.~J., Harris, H.~C., Pier, J.~R., Luginbuhl, C.~B., \& Ables, 
H.~D.\ 1992, \aj, 103, 638 
\bibitem[Monet(1998)]{usnoa2} Monet, D.~G.\ 1998, American 
Astronomical Society Meeting, 193, 112003 
\bibitem[Murison(2001)]{mur} Murison, M.\ 2001, 
\url{http://arnold.usno.navy.mil/murison/FAME/\\ObservationDensity/AstrometricErrors4.html}
\bibitem[Olling(2001)]{oll} Olling, R.~P.\ 2001, USNO Memorandum FTM2001-14,\\
 \url{http://ad.usno.navy.mil/\~{}olling/FAME/FTM2001-14.ps}
\bibitem[Olling \& Merrifield(2001)]{2001MNRAS.326..164O} Olling, R.~P.~\& 
Merrifield, M.~R.\ 2001, \mnras, 326, 164 
\bibitem[Olling \& Merrifield(1998)]{1998MNRAS.297..943O} Olling, R.~P.~\& 
Merrifield, M.~R.\ 1998, \mnras, 297, 943 
\bibitem[Reid, Readhead, Vermeulen, \& Treuhaft(1999)]{reid} 
Reid, M.~J., Readhead, A.~C.~S., Vermeulen, R.~C., \& Treuhaft, R.~N.\ 
1999, \apj, 524, 816 
\bibitem[Richards et al.(2001)]{rich} Richards, G.~T.~et 
al.\ 2001, \aj, 122, 1151 
\bibitem[Salim \& Gould(1999)]{R_0} Salim, S.~\& Gould, A.\ 
1999, \apj, 523, 633 
\bibitem[Scholz, Meusinger, \& Jahreiss(2001)]{scholz} 
Scholz, R.-D., Meusinger, H., \& Jahreiss, H.\ 2001, \aap, 374, L12 
\bibitem[Silvestri et al.(2001)]{silvestri} Silvestri, N.~M., 
Oswalt, T.~D., Wood, M.~A., Smith, J.~A., Reid, I.~N., \& Sion, E.~M.\ 
2001, \aj, 121, 503
\bibitem[USNO(1999)]{fame} USNO 1999, FAME Concept Study Report, USNO: 
Washington, DC \\(available at \url{ http://www.usno.navy.mil/FAME/publications})
\bibitem[Veron-Cetty \& Veron(2000)]{veron} Veron-Cetty, M.~P.~\& Veron, P.\ 2000, ESO Scientific Report
\bibitem[White et al.(2000)]{fbqs} White, R.~L.~et al.\ 
2000, \apjs, 126, 133 
\bibitem[Wilkinson \& Evans(1999)]{wilk} Wilkinson, M.~I.~\& 
Evans, N.~W.\ 1999, \mnras, 310, 645 
\bibitem[Yanny et al.(2000)]{yanny} Yanny, B.~et al.\ 2000, 
\apj, 540, 825
\bibitem[York et al.(2000)]{sdss} York, D.~G.~et al.\ 2000, 
\aj, 120, 1579 
\bibitem[Zheng et al.(2001)]{zheng} Zheng, Z., Flynn, C., 
Gould, A., Bahcall, J.~N., \& Salim, S.\ 2001, \apj, 555, 393 

\end{thebibliography}
\end{document}